\documentstyle[14pt,amssymb,a4]{article}
\begin{document}
\begin{center}
{\large {\bf Revision of upper estimate of percolation threshold on
square lattice}}
\end{center}

\begin{center}
{\bf Yu.P. Virchenko, Yu.A. Tolmacheva}
\end{center}

{\small
The more exact upper estimate of the percolation threshold
for the {\it site problem} on the quadratic lattice ${\Bbb Z}^2$ have been
found on the basis of the cluster decomposition.
It is done by the number estimate of cycles on ${\Bbb Z}^2$ which
maybe external boundaries of finite clusters.
\smallskip

\begin{center}
{\bf 1.\ Introduction}
\end{center}

Generally, percolation theory studies the connectedness
relation for random sets in topological spaces. Problems representing
greatest interest in the theory appear in
noncompact topological spaces.

The existence of a noncompact connected component in random set realizations
with nonzero probability is the crucial question in the theory [1].
In such a general setting of the problem, however, it is impossible to
obtain any strong results. That is why the basic objects of the percolation
theory are random sets in ${\Bbb R}^d$ and ${\Bbb Z}^d$ at $d=2,3$
generated by stationary random fields [1, 2].
The first case corresponds to so-called {\it continuous percolation theory}
and the second one -- to {\it discrete percolation theory}.
This limitation is a consequence of the fact that it is the problem
set in such a way is mainly needed in physical applications.
Therefore, it may be considered as the object of mathematical physics.
However, despite the above restriction, the basic problem
is very complicated and hardly lends itself to
exact mathematical study. Rigorous results referring to
discrete percolation theory which is its mostly developed direction
have been summarized in the monograph [3]. Later
results and also results referring to continuous theory have been summarized
in the excellent review [4]. Physical literature dealing with different
nonrigorous heuristic approaches to the investigation
of the percolation phenomenon, computer experiments and special
applications of the theory in theoretical physics is giant
and we shall not pay attention to it in this short introduction
to the problem.

Despite of the great flow of publications at 80-th in physical literature
and simultaneously appearance of some reviews in which mathematical results
are summarized, main questions of the percolation theory considered as the
object of mathematical physics remain without answers up to now.
It is the situation even for simplest case being studied in the theory
when random set is generated by Bernuolli's field on ${\Bbb Z}^2$.
Even for this case any algorithm of the {\it percolation threshold}
calculation (see \S 2) and, more generally, any algorithm of
percolation probability evaluation was not found for an arbitrary
{\it periodic graph} [1].
Our communication is dealing with the revision of upper estimate of
percolation threshold as compared to what may be found in literature
[4, 5] for the periodic graph on ${\Bbb Z}^2$ which
is called the {\it square lattice}.

\begin{center}
{\bf 2.\ Percolation theory problem on ${\Bbb Z}^2$}
\end{center}

Let us consider an infinite graph with the vertex set ${\Bbb Z}^2$.
For simplicity of statements and arguments, we shall study it as one
immersed into ${\Bbb R}^2$. The adjacency relation on the graph is defined
by the set of pairs $\{(x,y)\in {\Bbb Z}^{2}: x\varphi y\}$ where
$x\varphi y$ if and only if $y=x\pm {\bf e}_1$ or $y= x\pm {\bf e}_2,
{\bf e}_1=(1,0), {\bf e}_2=(0,1)$.
We shall follow the terminology of statistical physics and that is why
we shall call such a graph the {\it square lattice} and denote it by
the same symbol ${\Bbb Z}^2$.

Let $\{ \tilde c(x)\}$ is the Bernoulli random field
with the parameter
$$
c\ = \ {\rm Pr}\{{\tilde c}(x)=1\}
$$
which is said to be the concentration.
Below, the tilde marks the randomness of used objects.
The field $\{\tilde c(x)\}$ induces the random set with realizations
$\{ \tilde M: \tilde M\subset {\Bbb Z}^2 \}$ where
$\tilde M = \{ x:\tilde c(x)=1\}$. Sometimes we shall call
these realizations
{\it configurations of completed vertices} or, simply,
{\it configurations}.
This probability distribution for possible realizations
$\tilde M$ is clearly determined completely by the following probability
collection
$$
{\rm Pr} \{ \tilde M: A\subset \tilde M\}
\ =\ c^{|A|}\,, \qquad A\subset {\Bbb Z}^2\,.
$$
Here and further $|\cdot|={\rm Card}\{\cdot\}$.

Naturally, the adjacency relation $\varphi$ induces the connectedness
relation for vertices having included into any configuration $\tilde M$.
We shall call  two vertices $x$ and $y$ linked in $\tilde M$ if there
exists a path
$(x_i;i=0,1,2,...,n)$, $x_i \in \tilde M$, $x_0=x, x_n=y$ and
$x_i\varphi x_{i+1}$, $i=0,1,2,...,n-1$. The connectedness of vertices
of ${\tilde M}$ is the equivalence relation.
Therefore, each configuration $\tilde M$ is uniquely decomposed on some
disjoint equivalence classes
$\tilde M = \displaystyle\bigcup\limits_{j\in {\Bbb N}}{\tilde W_j}$
generated by the connectedness relation of vertices.
These classes we shall name {\it clusters}.
The collection of clusters related to configurations $\tilde M$
will be denoted by ${\cal W}[\tilde M]\ =\ \{\tilde W_j; j \in {\Bbb N}\}$.
If $x\in \tilde W_j$ for a $j \in {\Bbb N}$ in configuration $\tilde M$, then
we shall denote this cluster $\tilde W_j$ by $\tilde W(x)$ [3].

Let us introduce the random field $\{\tilde a(x); x\in {\Bbb Z}^2\}$
on the base of the set $\tilde M$,
$$
{\tilde a}(x)\ =\ \cases{1\,; & $x \in {\tilde W} \subset {\tilde M},\
|{\tilde W}(x)| < \infty$\,,\cr 0\,; & otherwise\,.\cr}
$$
In view that $\{\tilde c(x)\}$ is the uniform field (i.e.
the probability measure is invariant relative to translations on vectors
$(n_1 {\bf e}_1 + n_2 {\bf e}_2)$, $n_1, n_2 \in
{\Bbb Z}$), the random field $\{\tilde a(x)\}$
is uniform too. Therefore, the probability
$$
Q(c)\ =\ {\rm Pr}\{\tilde a(x)\ =\ 1\}
$$
does not depend on $x\in {\Bbb Z}^2$. This probability is a nondecreasing
function of $c$ [3]. If $Q(c) > 0$, then the {\it percolation} on
the $\{ \tilde c(x)\}$ is said to exist.
In connection with this fact, the following characteristic value is
introduced
$$
c^*\ =\ \sup \{c:\,Q(c)\ =\ 0\}\,.
$$
$c^*$ is called the {\it percolation threshold}.

\begin{center}
{\bf 3.\ Finite clusters on ${\Bbb Z}^2$}
\end{center}

At construction of some upper estimates of the probability $Q(c)$, it is
necessary to enumerate all finite clusters $W(x)$ including the fixed
vertex $x$.
We perform the enumeration by using the concept of the {\it external boundary}
for each finite cluster.
To this end, let us introduce on ${\Bbb Z}^2$ the following new concept of
the adjacency relation $\bar \varphi$ side by side
with the adjacency relation $\varphi$ [3,4].
We denote $\overline {{\Bbb Z}}^2$ the corresponding graph having the same
vertex set but with the adjacency relation $\bar \varphi$.

Vertices $x$ and $y$ are named the $\bar \varphi$-adjacent ones if
one of two cases takes place: 1) $x\varphi y$, 2)
either $y=x+{\bf e}_1\pm {\bf e}_2$ or $y=x-{\bf e}_1\pm {\bf e}_2$.

The relation $\bar \varphi$ induces the new relation of vertex connectedness
on each configuration $\tilde M$. This connectedness relation is the
equivalence relation too and it leads to a decomposition of $\tilde M$
on some connected sets of vertices relative to $\bar \varphi$.

Let us introduce the external boundary concept of a finite cluster on
${\Bbb Z}^2$.

D e f i n i t i o n\ 1.\ {\it The set $\partial W$ is named the boundary of
the cluster  $W$ if it consists of those vertices $y$ which
are $\varphi$-adjacent to vertex $x$ in the cluster $W$ but do not belong
to it.}

D e f i n i t i o n\ 2.\ {\it The external boundary $\bar \partial W$ of the
cluster $W$ is the set of vertices  $u\in \partial W$
such that for each of them there exists an infinite $\varphi$-path
$\alpha (u)$ on completed vertices of ${\Bbb Z}^2$ and,  moreover,
$u$ is the unique vertex in $\alpha (u)$ belonging to the union
$W\bigcup \partial W$.}

Further classification of finite clusters $W(x)$ is performed by the
enumeration of all possible external boundaries $\bar \partial W(x)$.
Following statement is the key for such enumeration. It repeats the
corresponding statement in the monograph [3] with the exception of the
last item.

{\bf Theorem \ 1}.\ {\it Let $W(x)$ be a finite cluster for a fixed
vertex $x$, $|W(x)|<\infty$.
Then $W(x)$ has a nonempty finite external boundary
$\bar\partial W(x)$ having following properties.

{\rm 1}.\ $\bar\partial W(x)$ is the $\bar \varphi$-connected vertex set
in ${\bar {\Bbb Z}}^2$ which presents the {\it cycle} i.e.
$\bar\partial W(x)\ =\ (x_1,x_2,...,x_n)$ where $x_i\not = x_j, \,
i\not = j, \,
n\ =\ |\bar\partial W(x)|, \, x_i\bar \varphi x_{i+1}, \,
i\ =\ 1,2,...,n$, $x_{n+1} = x_{0}$ and each vertex has only two
${\bar \varphi}$-adjacent vertices in the $\gamma$.
It is possible to introduce the definite orientation
on $\bar\partial W(x)$. (it will be used the
counter-clockwise orientation in further arguments.)

{\rm 2}.\ The vertex $x$ is contained in the finite set
${\rm Int}[\bar\partial W(x)]$ defined by
$${\rm Int}[\bar\partial W(x)]\ \equiv\
\{u \not \in \bar\partial W(x)\,:\,
\forall (\alpha (u): |\alpha (u)|=\infty)\,
\left( \alpha (u)\cap \bar\partial W(x)\not = \varnothing\right)\}\,;
$$

{\rm 3}.\ Let $u,v,w \in \bar\partial W(x)$ be three
$\bar \varphi$-adjacent vertices following one after another according to
the introduced orientation. Then if $u\varphi v$, the vertex $w$
belongs necessarily to one of the following collection:

{\rm a)}\ in the case $u\varphi v$ the set of three elements
$\{2v-u,\, 2v-u\pm {\bf e}\}$ where ${\bf e}$ is a unit basis vector
that is orthogonal to the vector $(v-u)$;

{\rm b)}\ in the case $v  = u + \varepsilon({\bf e}_1 - {\bf e}_2)$
the set of five elements
$$\{2v-u, \, 2v- u + \varepsilon {\bf e}_1,\, 2v - u + \varepsilon
{\bf e}_2,\, 2v - u \pm {\bf e}'/2\}$$
in the case $v  = u + \varepsilon({\bf e}_1 + {\bf e}_2)$\,;
$$\{2v-u, \, 2v - u + \varepsilon{\bf e}_1,\, 2v - u - \varepsilon
{\bf e}_2,\, 2v - u \pm {\bf e}'/2\}$$
where ${\bf e}'$ be an orthogonal vector to $(v-u)$ and
$|{\bf e}'|=\sqrt {2}$ and $\varepsilon = \pm 1$.}

The proof of first and second statements of Theorem 1 is obvious (see [3]).
What about the last statement, it is easy to verify its justification
from the explaining figure 1(a, b). The full rigorous proof is found
by us; it is quite tedious and goes beyond the scope of this communication.
Unlike the proof of the analogous theorem in [3], our arguments
are not used an application of Jordan's theorem.
Further we omits the proof and pass to the proof of the main statement
in next paragraphs.

\begin{center}
{\bf 4.\ The cluster decomposition on ${\Bbb Z}^2$}
\end{center}

We shall consider the probability $1 - Q(c) = {\rm Pr}\{\tilde a(0)=0\}$. Let
${\cal A} = \{W: W = W(0)  \mbox{ is cluster}\,, |W|< \infty\}$ be
the collection  of all finite clusters including the vertex 0.
We define the event
$$
A (W)\ =\ \{\tilde M: \tilde W=W, \, 0\in \tilde M, \,
\tilde W\in \{ \tilde W_j; j \in {\Bbb N}\}\}
$$
for any cluster $W\in {\cal A}$.
This event has the definite probability
$$
{\rm Pr}\{A(W)\}\ =\ c^{|W|}(1-c)^{|\partial W|}\,. \eqno(1)
$$
According to statements in the previous section,
any cluster of this collection ${\cal A}$ is corresponded to a cycle
$\gamma$. Vertices of this cycle are $\bar \varphi$-adjacent and such that
$0\in {\rm Int}[\gamma]$. In this connection, let us introduce in
consideration the collection $\cal B$ of all $\bar \varphi$-cycles  having
the last property. Let $B(\gamma)$ be the event
$$
B(\gamma)\ =\ \{ \tilde M\,:\, 0\in \tilde M, \tilde W(0)\in
{\cal W}[\tilde M], \bar\partial\tilde W(0) = \gamma\} \eqno(2)
$$
defined for any $\bar \varphi$-cycle $\gamma \in \cal B$. It
is represented by the finite union of mutually disjoint events
$$
B(\gamma)\ =\ \bigcup^{}_{W\in {\cal A}\,:\,
\bar\partial W = \gamma} A(W)\,. \eqno(3)
$$
Hence, according to (1) and (3), such event has the definite probability
$$
P(\gamma)\ =\ {\rm Pr}\{B(\gamma)\}\,.
$$
It is equal
$$
P(\gamma)\ =\ \sum^{}_{W\in {\cal A}\,:\,\bar\partial W =\gamma}
{{\rm Pr}\{A (W)\}}\ =\ \sum^{}_{W\in {\cal A}\,:\,\bar\partial W =\gamma}
c^{|W|}(1-c)^{|\partial W|}\,.
$$
Let us note that
$$
\{\tilde a(0)\ =\ 0\}\ =\ \bigcup^{}_{W\in {\cal A}} A(W)\,. \eqno(4)
$$
The collection ${\cal A}$ is decomposed on some disjoint classes of
clusters. Each class consists of those clusters $W\in {\cal A}$ which have
the same external boundary. It is realized $\bar\partial W =\gamma$ for them.
Therefore, the following representation is correct
$$
\bigcup^{}_{W\in {\cal A}}{...}\ =\
\bigcup^{}_{\gamma\in {\cal B}}\ \
{\bigcup^{}_{W\in {\cal A}\,:\,\bar\partial W =\gamma}{...}}\,.
$$
Then we obtain
$$
\{\tilde a(x)\ =\ 0\}\ =\ \bigcup^{}_{\gamma \in {\cal B}}B(\gamma)
$$
using (3). Finally, we come to the statement

{\bf Theorem\ 2}.\ {\it The probability $1 - Q(c)$ is represented by the
decomposition}
$$
1 - Q(c)\ = \ \sum^{}_{\gamma\in {\cal B}}P(\gamma)\,.
\eqno(5)
$$

Usually, such decomposition is said to be the {\it cluster} one
in the percolation theory [4].

\begin{center}
{\bf 5.\ The main theorem}
\end{center}

The cluster decomposition (5) is represented by the sum of
probabilities of some disjoint events. Therefore, definitely, the cluster
decomposition is convergent.
The function $Q(c)$ is not equal to zero only if $c > c^* > 0$.
So, it is not an analytic function on the concentration $c$
and $c = c^*$ is its the singular point.
Earlier [4], it was obtained that $6/7 > á^* > 1/3$.
We give the improvement of the upper estimate at the below-formulated
statement.

{\bf Theorem \ 3}.\ {\it The inequality $c^*\leq c_0 = 3-\sqrt {5}$ is
correct for Bernoulli's random field on ${\Bbb Z}^2$.}

P r o o f\ .\ We use the elementary estimate
$$
P(\gamma)\leq (1\ -\ c)^{|\gamma|}\,,
$$
that follows from the Definition 2 and Eq.(2).
Using it and Eq.(5), we come to the upper boundary
$$
\sum^{}_{\gamma \in {\cal B}}P(\gamma)\ \leq\
\sum^{}_{\gamma \in {\cal B}}{(1\ -\ c)^{|\gamma|}}
\ =\ \sum^{}_{\gamma \in {\cal B}\,: |\gamma | = 2k}
{(1\ -\ c)^{2k}r_k} \eqno (6)
$$
where $r_k= {\rm Card}\{\gamma\in {\cal B}\,:\,|\gamma| = 2k\}$, $k\geq 2$.

Further, we shall find the upper estimate for the value $r_k$.
Consider the infinite path $\alpha(0)=(j{\bf e}_1;\, j=0,1,2...)$
with initial vertex $0$. Then, according to Theorem 1 (2),
each cycle $\gamma\in {\cal B}$ necessarily crosses the path in some vertices.
Let us select the vertex in this set of all intersection
vertices which is nearest to the vertex $0$. Denote it by $z_{\gamma}$.
All cycles of ${\cal B}$ is decomposed on disjoint classes ${\cal C}_\gamma$,
i.e.
the cycles having the same vertex $z_{\gamma}$ belong to the same class.
This class decomposition induces the decomposition of the cycle set
$\{\gamma \in {\cal B}:|\gamma| = 2k\}$ on corresponding
classes ${\cal C}_l^{(k)}$  where $l$ is the distance from 0 to $z_\gamma$.
In addition, $l < k$ and therefore,
$$
\{\gamma\in {\cal B}\,:\,|\gamma| = 2k\}\ =\ \bigcup_{l=1}^{k-1}
{\cal C}^{(k)}_l\,,\qquad r_k\ =\ \sum_{l=1}^{k-1} |\,{\cal C}^{(k)}_l|\,.
$$
Let $\gamma = (z_\gamma = x_0, x_1, ..., x_{2k-1}, x_{2k} = z_\gamma)$.
According to  the earlier introduced orientation, the vertex $x_1$ that
follows after  $x_0$ in the cycle $\gamma$ may be one of the set only
$$
\left(z_\gamma + {\bf e}_1, z_\gamma + {\bf e}_1 + {\bf e}_2,
z_\gamma + {\bf e}_2, z_\gamma - {\bf e}_1 + {\bf e}_2\right)\,. \eqno (7)
$$
Using this ordering, cycles belonging to the same class
${\cal C}^{(k)}_l $ are distributed on disjoint collections
${\cal C}^{(k,i)}_l $, $i = 1, 2, 3, 4$
independence of the selection of the vertex $x_1$ in the cycle $\gamma$.
Hence,
$$
{\cal C}^{(k)}_l\ =\ \bigcup_{i=1,2, 3, 4} {\cal C}^{(k,i)}_l\,,\qquad
\left|\,{\cal C}^{(k)}_l\right|\ =\
\sum_{i =1, 2, 3, 4} \left|\,{\cal C}^{(k,i)}_l\right|\,.
$$
We obtain as the result
$$
r_k\ =\ \sum_{l=1}^{k-1}
\sum_{i =1, 2, 3, 4} \left|\,{\cal C}^{(k,i)}_l\right|\,.
$$
We must find upper estimate of the value $\left|\,{\cal C}^{(k,i)}_l\right|$.
It is easy to see that ${\cal C}^{(k,i)}_l\subset {\cal P}^{(2k-1,i)}_l$
where  ${\cal P}^{(2k-1,i)}_l$ is the set of paths beginning at the vertex
$z_\gamma$ and the distance from $z_\gamma$ to the vertex 0 along
the path $\alpha (0)$ is equal to $l$. Besides, they have the length
$2 k -1$ and the vertex $x_1$ is the $i$-th in the finite sequence (7).
We sort out paths $\gamma$ with the length $2k-1$ but
not with $2k$. It is connected with the fact that the last edge
$x_{2k-1} {\bar \varphi} x_0$ of the cycle $(x_0, x_1, ..., x_{2k-1}, x_0)$
is fixed by the collection of vertices $x_0, x_1, ..., x_{2k-1}$.
Therefore,
$$
\left|\,{\cal C}^{(k,i)}_l\right|\ \le\ \left|\,{\cal
P}^{(2k-1,i)}_l\right|\ =\ s_{2k-1}\,.
$$
It is easy to see that $s_n$ does not depend on $l$ and $i$ due to our
construction. Let us represent
$$
s_n\ =\ s_n^{+} \ +\ s_n^{\times} \eqno (8)
$$
where
$$
s_n^{+}\ =\ {\rm Card}\{\gamma\in {\cal P}^{(n,i)}_l\,:\,
x_{n-1} \varphi x_{n}\}\,,
$$
$$
s_n^{\times}\ =\ {\rm Card}\{\gamma\in {\cal P}^{(n,i)}_l:\, \overline{
x_{n-1} {\varphi\ } x_{n}},\ x_{n-1} \bar\varphi x_{n}\}.
$$
We in\-tro\-du\-ce the two-\-com\-pon\-ent vec\-tor
$\left[s_n^+, s_n^\times\right]$ (not distinguishing co\-lumn-\-vec\-tors
and tup\-le-\-vec\-tors).
Then the following equation takes place
$$
\left[\matrix{s_n^+ \cr s_n^\times\cr}\right]\ =\ {\sf T}
\left[\matrix{s_{n-1}^+ \cr s_{n-1}^\times\cr}\right]
$$
with the transfer matrix
$$
{\sf T}\ =\ \left[\matrix{1 & 2\cr 2 & 3\cr}\right].
$$
according to Theorem 1 (3).
Hence,
$$
\left[\matrix{s_n^+ \cr s_n^\times\cr}\right]\ =\ {\sf T^{n-1}}
\left[\matrix{s_{1}^+ \cr s_{1}^\times\cr}\right]\,. \eqno (9)
$$
Eigenvalues of the matrix $\sf T$ is equal to
$\lambda_+=2+\sqrt {5},\, \lambda_- = 2 - \sqrt {5}$
and corresponding eigenvectors (they are orthogonal but are not normalized)
have the form
$$
{\sf t}_+\  =\ \left[\matrix{1 \cr \frac{1+\sqrt {5}}{2}\cr}\right]\,,
\qquad
{\sf t}_-\  =\ \left[\matrix{1 \cr \frac{1-\sqrt {5}}{2}\cr}\right]\,.
$$
Decomposing the vector $\left[s_1^+ , s_1^\times\right] = [2, 2]$ on
eigenvectors ${\sf t}_+, {\sf t}_-$,
$$
[s_1^+, s_1^\times]\ =\ g_+ {\sf t}_+\ +\ g_- {\sf t}_-\,,
$$
we obtain $g_+  = 1 + \sqrt {5}/{5}$, $g_- = 1- \sqrt {5}/{5}$.
Therefore,
$[s_n^+, s_n^\times] = g_+ \lambda_+^{n-1} {\sf t}_+ + g_- \lambda_-
{\sf t}_-$ and, according to (8), (9), we find
$$
s_n\ =\ \left(g_+ \lambda_+^{n-1} + g_-\lambda_-^{n-1}\right) +
\frac 12\left(g_+\lambda_+^{n-1} (1 + \sqrt{5}) + g_- \lambda_-^{n-1}
(1 - \sqrt{5})\right)\ \le
$$
$$
\leq 4\left(2 + \sqrt {5}\right)^{n-1}\,. \eqno (10)
$$
Since $r_k < 4(k-1)s_{2k-1}$, hence, putting $n = 2k-1$
at the estimate having found, we obtain the majorant series from (6).
It is summable at $c > 3 - \sqrt {5}$.
Then, if this condition is satisfied, we shall obtain
$$
\sum^{}_{\gamma\in {\cal B}}{P(\gamma)}\ \leq\ \infty\,. \eqno (11)
$$

Completion of the proof is performed on the base of arguments which are
standard in the percolation theory (see, for example, [5]).
The inequality (11) permits to apply the Bor\-el-Can\-tel\-li lem\-ma
(see, for example, [6]) to the event family ${\cal B}$. According to
this statement, the probability of the event that consists in simultaneous
realization of an infinite set of events belonging to the family, is equal to
zero. Then, with the probability one, there exists a maximal
cycle $\gamma\in {\cal B}$.
For this cycle, there is a vertex $z$ out of it and, besides,
there is the infinite path $\alpha (z)$ without intersections and with the
initial vertex $z$. The path does not intersect the cycle.
This  means that the event
$\{\tilde M:\, \exists \left(\tilde W \in {\cal W}[\tilde M]\right)
\left(|\tilde W| = \infty\right)\}$ of Bernoulli's field $\{{\tilde c}(x)\}$
has the probability 1. On the other hand, the countable
decomposition
$$
\{\tilde M:\, \exists \left(\tilde W \in {\cal W}[\tilde M]\right)
\left(|\tilde W| = \infty\right) \} = \bigcup_{v
\in {\Bbb Z}^2}\{{\tilde M}:\,\tilde W(v) \in {\cal W}[{\tilde M}],
|\tilde W(v)| = \infty\} \eqno (12)
$$
takes place. In view of the random field $\{{\tilde a}(x)\}$ is uniform,
the probability
$$
{\rm Pr}\{{\tilde M}:\,\tilde W(v) \in {\cal W}[{\tilde M}],
|\tilde W(v)| = \infty\} = Q(c)
$$
does not depend on $v$.
Therefore, it cannot be equal to zero, since the following inequality is
correct according to (12)
$$
1\ \le\ \sum_{v \in {\Bbb Z}^2}{\rm Pr}
\{{\tilde M}\,:\,\tilde W(v) \in {\cal W}[{\tilde M}],
|\tilde W(v)| = \infty\}\,. \qquad \blacksquare
$$

{\bf Consequence}.\ {\it At $á > c_0$, the probability $Q(c)$ may be
represented with any preassigned accuracy by finite sum of series} (5).
{\it It is defined by summands which correspond to cycles $\gamma$ with the
length not exceeding} $2m$ with fixed $m\in \Bbb N$.
{\it In this case, the following error estimate takes place}
$$
\sum_{\gamma\in {\cal B}: |\gamma | > 2m}{P(\gamma)}\ \le\
\sum_{\gamma\in {\cal B}: |\gamma | > 2m}{(1-c)^{|\gamma|}}\le
\sum^{}_{k > m}(1-c)^{2k-1}s_ {2k-1}\ =
$$
$$=\ 16(1-c)^{2}
\sum^{\infty}_{k = m+1}{(k-1)\left[\left(2 +\sqrt{5}\right)(1-c)
\right]^{2k-2}}\ =
$$
$$
=\ 16(1-c)^{2}\xi
\left[ \frac{d}{d\xi}\left(\frac{\xi^m}{1-\xi}\right)\right]_
{\xi=\left[\left(2 +\sqrt{5}\right)(1-c)\right]^2 } <\ \infty \qquad
\blacksquare
$$

Thus, the decomposition (5) is the key for solving of the main
percolation problem on ${\Bbb Z}^2$
that is to calculate the probability $Q(c)$
with any guaranteed
accuracy.
In addition, due to presence of the singular point $á = c^*$, the majorant
series for the remainder must have a singularity in a point $á_0 > c^*$
for any
initial finite sum of the cluster decomposition. Such a situation will take
place in any way of upper estimation of the percolation threshold.
Then, it
follows that the solution of the main problem is impossible without
building of a
calculation algorithm  for the percolation threshold $c^*$ with any
preassigned accuracy.

\begin{center}
{\bf 6.\ Discussion}
\end{center}

One can see on the base of the above proof that the singularity
$á^*$ is possibly connected with the singularity of the generating
function for the nonintersecting contours number. If it is true, so the
regular method of sequential approximative calculation of this point
may be based on the sequential exclusion of intersections in paths
$(z_\gamma, x_1, ..., x_{2k-1})$ corresponding to cycles $\gamma$.
Such an exclusion must be fulfilled with the increasing of approximation
order.
In this case the estimate given by Theorem 2 may be considered as
the zero approximation.

Generally, the algorithm of approximations of the
point $c^*$ should be consist in such a construction when estimates
$á_n^- < c^* < c_n^*$, $c_{n+1}^- > c_n^-$, $c_{n+1}^+ < c_n^+$ are
build sequentially step by step. They have the property
$\lim\limits_{n \to \infty} c_n ^{\pm} = c^*$ and besides, for sure,
$Q(c) = 0$ at $c < c_n^-$; $Q(c) > 0$ at $c > c_n^+$.
\eject
\thispagestyle{empty}
\begin{center}
{\large {\bf References}}
\end{center}

\noindent
1. {\it Yu.P. Virchenko}, {\bf Percolation.
Mathematical Physics. Great Russian Encyclopedia. Moscow (1998),
p.346-347  (in Russian)}
\medskip

\noindent
2. {\it Yu.P. Virchenko}, {\bf Per\-co\-la\-ti\-on of ran\-dom fi\-eld.
Ma\-the\-ma\-ti\-cal Phy\-sics. Great Russian Encyclopedia. Moscow (1998),
p.363-364  (in Russian)}
\medskip

\noindent
3. {\it ~H. Kesten}, {\bf Per\-co\-la\-tion The\-ory for
Ma\-the\-ma\-ti\-ci\-ans. Boston. Birkh{\" a}user (1982)}
\medskip

\noindent
4. {\it M.V. Men'shikov, S.A. Molchanov, A.F. Sidorenko},
{\bf Percolation Theory
and Some Its Applications. --  In: Advances in Sci. and Engin., Ser.
Propability Theory, Mathematical Statistics and Theoretical
Cybernetics , Vol. 24. Moscow (1986), VINITI Publ. 53p. (in Russian)}
\medskip

\noindent
5. {\it V.A. Malyshev, M.V.  Men'shikov, E.V. Petrova}, {\bf Introduction in
Probability Theory.
Moscow (1997), MGU Publ, p.119  (in Russian)}
\medskip

\noindent
6. {\it J. Lamperti}, {\bf Probability. New York - Amsterdam (1966),
Dartmuoth college}
\medskip

\end{document}